\documentstyle[preprint,pre,aps,epsf,graphicx]{revtex}
\begin{document}

\draft

\title{A Model for Phase Transition based on Statistical Disassembly of Nuclei
at Intermediate Energies}
 
\author{G. Chaudhuri, S. Das Gupta, and M. Sutton}

\address{Physics Department, McGill University, 
Montr{\'e}al, Canada H3A 2T8}

\date{\today}

\maketitle

\begin{abstract}
Consider a model of particles(nucleons) which has a two-body interaction
which leads to bound composites with saturation properties.  These
properties are : all composites have the same density and the 
ground state energies
of composites with $k$ nucleons are given by $-kW+\sigma k^{2/3}$
where $W$ and $\sigma$ are positive constants.  $W$ represents
a volume term and $\sigma$ a surface tension term.  These 
values are taken from nuclear physics.  We show that in the large $N$
limit where $N$ is the number of particles
such an assembly in a large enclosure at finite temperature
shows properties of liquid-gas phase transition.  We do not use the
two-body interaction but the gross properties of the composites only.
We show that (a) the $p-\rho$ isotherms show a region where pressure
does not change as $\rho$ changes just as in Maxwell construction
of a Van der Waals gas, (b) in this region the chemical potential
does not change and (c) the model obeys the celebrated Clausius-Clapeyron
relations.  A scaling law for the yields of composites emerges.

For a finite number of particles $N$ (upto some thousands)
the problem can be easily solved on a computer.
This allows us to study finite particle number effects which modify
phase transition effects.

The model is calculationally simple.  Monte-Carlo simulations are not
needed.

\end{abstract}


\section{Introduction}

A very popular and highly successful model for collisions of two nuclei
at intermediate energies (50 MeV to 100 MeV per nucleon) 
is the following.  Because of many collisions
between nucleons, a statistical equilibrium is reached.  
The temperature rises.  The system
expands from normal density and composites are formed on the way to
disassembly.  As the system reaches between 3 to 6 times the normal 
volume, the interactions between the composites become unimportant
(except for the long range Coulomb interaction) and one can do a 
statistical equilibrium calculation to obtain the yields of the
composites at a volume which is called the freeze-out volume.
Although the model is simple, actual realistic calculations
based on the model are much harder.  The nucleus is a finite system.
It has two kinds of particles, neutrons and protons (generically termed
nucleons).  Protons carry charges and prevent large nuclei from being
formed.  For realistic treatment, the idea of a strict freeze-out volume
has to be modified.

Here we consider the same physics but with the following 
simplifications: only one kind of particle is considered and the
Coulomb interaction is neglected meaning arbitrarily large ``nuclei'' can be
formed.
The energy scale is MeV (million electron volt) and the
length scale is fm ($10^{-13}$cm) so
the salient features of nuclear physics are retained.
The binding energy and the volume of a composite is proportional
to the number of particles (nucleons) in the composite and
have a surface tension proportional to the surface area.  We
show, with rather little effort, that the model leads to a first order
phase transition as either the density the temperature or both are varied.  The
system has a region of liquid-gas coexistence where, as for the Maxwell
construction of a Van der Waals gas, pressure remains constant when
the density increases along the isotherm. 
In this region the chemical potential remains unchanged.  As one
traverses the path from the liquid phase to the gas phase the 
Clausius-Clapeyron relationship is obeyed.  For large systems,
a scaling law for composites emerges: if we know the yields of composites
for one large system, we know these for another large system.

A more realistic version of this model has been used for Bevalac physics 
($>$250 MeV per nucleon beam energy in the lab) by many
authors more than twenty-five years ago.  It is not possible to
quote all the references but a review article \cite{Dasgupta1}
has a
more complete list.
The possibility of a phase transition was not considered as
the collision energies were too high for the
liquid phase and only very light composites could be produced.

Phase transitions in heavy ion collisions at intermediate energies
became a topic of considerable interest starting from the mid-eighties
and continues to be a central issue.  There are many approaches
and a large literature too numerous to list.  We will refer here
to only a few which closely follow the underlying physics considered
here.  The same model as used here was adopted in \cite{Dasgupta2} for
finite nuclei.  By extrapolation it was shown that the model
leads to a first order phase transition.  A brief 
application of this model is given in section VII.  A grand canonical model
was used in \cite{Bugaev} which demonstrated a first order phase 
transition.  The approach was quite different from
what is used here.  We use simpler, more traditional and
numerical methods.
Our results are similar but sufficiently different
to warrant a full description.  A discussion of Clausius-Clapeyron 
relations and a scaling law highlight some interesting physics.

The celebrated statistical
Multifragmentation model (SMM) of Copenhagen \cite{Bondorf}, the 
microcanonical models of Gross and Randrup, Koonin \cite{Gross,Randrup}
use the same underlying physics as in this work.  But the emphasis
was on trying to be as close to to the actual nuclear situation as one can
and thus the phase transition aspects are largely hidden.

\section{The Basic Formulae}
If we have $n_a$ particles of type $a$, $n_b$ particles of type $b$, $n_c$
particles of type $c$ etc. all enclosed in a volume $V$ and interactions
between particles can be neglected, the grand partition function for
this case can be written as
\begin{eqnarray}
Z_{gr}=\prod_{i=a,b,c...}\exp(e^{\beta\mu_i}z_i)
\end{eqnarray}
Here the $\mu_i$ is the chemical potential and $z_i$  the 
canonical partition function of one particle of type $i$.  The average
number of particles of type $i$ is given by $\partial (ln Z_{gr})/
\partial (\beta\mu_i)$ :
\begin{eqnarray}  
n_i=e^{\beta\mu_i}z_i
\end{eqnarray}
It is possible that one of the species can be built from two other
species.  In reverse, a heavier species can also break up into two lighter
species.  If $\alpha$ number of particles of type $a$ can combine with
$\beta$ number of particles of type $b$ to produce $\gamma$ 
number of particles
of type $c$, then chemical equilibrium implies \cite{Reif} that the chemical
potentials of $a,b$ and $c$ are related by $\alpha\mu_a+\beta\mu_b=
\gamma\mu_c$.

In our model we have $N$ nucleons in a volume $V$ but these nucleons
can be singles or form bound dimers, trimers etc.  Chemical equilibrium implies
that a composite with $k$ bound nucleons has a chemical potential $k\mu$
where $\mu$ is the chemical potential of the monomer (nucleon).  Thus our
ensemble has monomers, dimers, trimers etc. upto some species with $k_{max}$
bound nucleons where ideally $k_{max}\rightarrow\infty$.  For practical
calculations, we use a finite value of $k_{max}$.  Most of the results shown
here use $k_{max}=2000$ although we have also done calculations
with much larger values.
Choosing $k_{max}=2000$ does not mean that the total number of nucleons
is 2000.  The total number of nucleons can be infinite but 
the largest species allowed in the calculation (is somewhat artificially)
limited to 2000. 
The total number of nucleons will be denoted by $N$ where $N$ is
very large. The quantity $k_{max}$ plays an essential role, setting
$k_{max}$ too low (for example 200 as shown
in section VII) makes then liquid-gas transition disappear.  An assembly
with $\simeq$200 particles or less does not display
the typical behaviour of liquid-gas co-existence.

We now look into $z_i$, the partition function of one composite of $i$
nucleons.This factors into two parts, a traditional translation energy part
and an intrinsic part:
$z_i=z_i(tran)z_i(int)$ where
\begin{eqnarray}
z_i(tran) & = &\frac{V}{h^3}\int \exp(-\beta p^2/2m_i)d^3p  \\
\nonumber
& = &\frac{V}{h^3}(2\pi m_iT)^{3/2}
\end{eqnarray}
The intrinsic part $z_i(int)$ of course contains the key to phase transition.
If we regard each composite to exist only in a ground
state with energy 
$e_i^{gr}$, then $z_i(int)=\exp(-\beta e_i^{gr})$.
We use $e_i^{gr}=-iW+\sigma i^{2/3}$ where nuclear physics sets $W$=16 MeV
and $\sigma=18$ Mev.  This simple model itself will
lead to the main results of this paper.  Because of the surface term,
energy per particle drops as $i$ grows.  Let us denote by $F$ the
free energy of the $N$ nucleons where $N$ is the total number of nucleons;
$E$ be the energy and $S$, the entropy: $F=E-TS$.  At finite temperature $F$
will go to its minimum value.  The key issue is how the system of $N$
nucleons breaks up into clusters of different sizes as the temperature
changes.  At low temperature $E$ and hence $F$ minimises
by forming very large clusters (liquid).  But as the temperature increases
$S$ will increase by forming larger number of clusters thus breaking up the
big clusters. Gaseous phase will appear.  How exactly this will  happen
requires calculation and we find that the system goes through a 
first order liquid-gas phase transition.

We used here a slightly more
sophisticated model for $z_i(int)$.  This does not make the calculation 
any harder (or alter the qualitative features) but
makes it more realistic.  We make the surface tension temperature
dependent in conformity with usual parametrisation \cite{Bondorf};
$\sigma(T)=\sigma_0[(T_c^2-T^2)/(T_c^2+T^2)]^{5/2}$

Here $\sigma_0=$18 MeV and $T_c$=18 MeV.  At $T=T_c$ surface tension 
vanishes and we have a fluid only.  For us this is unimportant as our
focus will be the temperature range 3 to 8 MeV.  

In computing the partition function $z_i(int)$ we include not just the ground
state but also excited states of the composite in an approximate 
fashion.  We should compute $z_i(int)=\exp(-\beta e_i^{gr})+
\int\sum g_i(e)\exp(-\beta e)$.  Here $e>e_i^{gr}$ and $g_i(e)$ is
the density of excited states of this particular composite.  Instead of
trying to calculate $z_i(int)$ by performing the sum and integral 
we use a well-known trick.  Utilise the relation 
$z_i(int)=exp(-f_i(int)/T)$
where $f_i(int)=e_i^T-Ts_i$ and now use the Fermi-gas formula for
the nucleus with $i$ nucleons (approximately correct and widely used
at intermediate temperature).  This gives $e_i^T=e_i^{gr}+iT^2/\epsilon_0$.
This is similar to electron gas at finite temperature
(excitation energy goes like $T^2$) 
except that in nuclear
physics the value of $\epsilon_0$ is $\simeq$16 MeV.  The intrinsic entropy 
of the
nucleus at this temperature is $2iT/\epsilon_0$.  The expression for 
$z_i(int)$ is now complete and easily tractable.

Let us now summarise the relevant equations.
For $k=1$ (the nucleon which has no excited states)
\begin{eqnarray}
n_1=\frac{V}{h^3}(2\pi mT)^{3/2}\exp(\mu/T)
\end{eqnarray}
and for $k>1$
\begin{eqnarray}
n_k=\frac{V}{h^3}(2\pi mT)^{3/2}k^{3/2}\exp[(\mu k+Wk+kT^2/\epsilon_0-
\sigma(T)k^{2/3})/T]
\end{eqnarray}
Here $n_k$ is the average number of composites with $k$ nucleons.  In the
rest of the paper, for brevity, we will omit the qualifier ``average''.
It is always implied.

A useful quantity is the multiplicity defined as
\begin{eqnarray}
M=\sum_{k=1}^{k_{max}}n_k
\end{eqnarray}
The number of nucleons bound in a composite with $k$ nucleons is
$kn_k$ and obviously $N=\sum_{k=1}^{k_{max}}kn_k$.
The pressure is given by
\begin{eqnarray}
p=\sum_{k=1}^{k_{max}}\frac{n_k}{V}T
\end{eqnarray} 
Quantities like $N,V,n_k$ are all extensive variables.  These
equations can all be cast in terms of intensive variables like $N/V=\rho,
n_k/N$ etc so that we can assume both $N$ and $V$ approach very large
values and fluctuations in the number of particles can be ignored.  Thus
for a given temperature and density we solve for $\mu$ using
\begin{eqnarray}
\rho=\frac{(2\pi mT)^{3/2}}{h^3}(\exp(\mu/T)+\sum_{k=2}^{k_{max}}k^{5/2}
\exp[(\mu k+Wk+kT^2/\epsilon_0-\sigma(T)k^{2/3})/T])
\end{eqnarray}
The sum rule $N=\sum_{k=1}^{k_{max}} kn_k$ changes to $1=\sum kn_k/N$.

From what we have described so far it would appear that $V$ in eqs.(3),
(4) and (6) is the freeze-out volume $V$, the volume to which the system
has expanded.  Actually if the freeze-out volume is $V$ then in these 
equations we use $\tilde V$ which is close to $V$ but less.  The reason 
for this is the following.  To a good approximation a composite of
$k$ nucleons is an incompressible sphere with volume $k/\rho_0$
where the value of $\rho_0$ is $\simeq$ 0.16 fm$^{-3}$.   The volume available
for translational motion (eq.(3)) is then $\tilde V=V-V_{excluded}$ where 
we approximate $V_{excluded}\simeq N/\rho_0=V_0$
the normal volume of a nucleus with $N$ nucleons.  Similar corrections
are implicit in Van der Waals equation of state.  This is meant
to take care of hard sphere interactions between different particles.
This answer is approximate.  The correct answer is multiplicity 
dependent.  The approximation of non-interacting composites in a volume
gets to be worse as the volume decreases.  We restrict our
calculation to volumes $V$ greater than $2V_0$.  This is how the calculations
reported in the next section proceed.  We choose a value of $V_0/V=
\rho/\rho_0$ from which 
$V_0/\tilde V=\tilde{\rho}/\rho_0=\rho/(\rho_0-\rho)$ is deduced.
This value of $\tilde{\rho}$ is used in eq.(8) to calculate $\mu$
and all other quantities.  We plot results as function of $\rho/\rho_0$.
If we plotted them as function of $\tilde{\rho}/\rho_0$ the plot would
shift to the right.

Calculations in \cite{Bugaev} are continued beyond the limit $\rho/\rho_0=0.5$.
They find one can identify a critical point at $T=T_c=18.0 MeV$,
$\rho/\rho_0=1$ and $p_c=\infty$.  At very high pressure the model
should break down: $z_k(int)$ must change at such high pressure although 
nuclear physics says that nuclei being highly incompressible moderate
pressures should leave the internal partition functions relatively unchanged.
Another way of saying this is that interaction between composites
should be taken into account for $V\simeq V_0$.

\section{The $p-\rho$ curves for isothermals}
For a given temperature $T$ and $\rho$ we solve for $\mu$ and then pressure.
This is plotted in Fig.~1.  For each isotherm shown the pressure rises rapidly 
at first with $\rho$ but then flattens out.  The flattening depends upon the 
value of $k_{max}$.  For low $k_{max}$ (shown in section VII) there will 
not be any flattening.  The value of $k_{max}$ used in Fig.~1  is 2000.
There is still a very slight rise in $p$ (not discernible in the Figure).
The figure empirically allows us to designate two regions:
a purely gas phase where the pressure rises with density and a liquid-gas 
co-existence phase where the volume changes but the pressure is nearly
stationary.  One way of seeing this is that
$p=T(M/\tilde{V})$.  As $\tilde{V}$ decreases
so does $M$ so as to compensate in the co-existence region.
Fig.~1 also shows that in the gas phase, the chemical potential rises rapidly
with density but then flattens out in the co-existence phase.  

A discussion about $\mu$ dependence in the co-existence region
is in order here.  For $k_{max}$ large
$\rho$ is much more sensitive to $\mu$ than pressure ($\rho$ has
weighting of $k^{5/2}$, eq.(8) whereas $p$ has $k^{3/2}$ weighting (eqs.
(5) and (7)).  For very large $k_{max}$ an infinitesimal change in $\mu$
will lead to a finite change in $\rho$ but only a very small change in $p$.
In the limit $k_{max}\rightarrow\infty$ we will reach ideal liquid gas 
phase transition: no change of $\mu$ in the coexistence region and no
change of pressure.  This is demonstrated in section VI.

\section{What constitutes the gas and what constitutes the liquid?}
As an example, at fixed temperature 7 MeV we show in Fig.~2 the distribution of
composites (a) in the pure gas region ($\rho/\rho_0=0.12$)) and (b) in the 
co-existence region ($\rho/\rho_0=0.22$). In the gas phase the sum rule
$\sum kn_k/N=1$ is exhausted well before we reach 50.  There are no heavy 
composites.  In the co-existence region there are light particles
($k\leq 40$), then nothing for a long range of $k$, and then there are heavy
particles with $k$ between 1800 to $k_{max}=2000$ (the figure
shows the population of $k$=1900 to $k_{max}=2000$).  A safe functional
definition for the gas phase is all composites between $k$=1 to
$k=100$ and for the liquid phase all composites between $k_{max}-300$
to $k_{max}$.  Thus both the liquid and the gas phases are quite
complicated; consisting of not one or two but many species although
they are all made up of the same elemental nucleon.

\section{The Clausius-Clapeyron relations}
In the co-existence region the pressure is a rapidly increasing function
of the temperature.  In Fig.~1 these are the flat regions shown for
three temperatures (dash-dot 6 MeV, solid 7 MeV and dash 7.5 MeV).
The Clausius-Clapeyron relation for liquid-gas phase transition 
provides an equation for the rate of change \cite{Reif}:
\begin{eqnarray}
\frac{dp}{dT}=\frac{\Delta s}{\Delta v}
\end{eqnarray}
where $\Delta s$ can be taken to be the change of entropy per unit mass
and $\Delta v$ the corresponding change in volume as matter moves across
phase transition.  We can take these changes to be per nucleon.  The following
substitutions are made:$\Delta s=L/T$ ($L$ is the latent heat) and 
$\Delta v=v_{gas}-v_{liq}$.  The standard approximation now is
 $v_{gas}>>v_{liq}, v_{gas}=1/\rho_{gas}\simeq T/p$
and thus 
\begin{eqnarray}
\frac{dp}{dT}\simeq \frac{Lp}{T^2}
\end{eqnarray}
If we make the assumption that $L$ is nearly independent of 
temperature then the equation integrates out to give
\begin{eqnarray}
ln~p=ln~p_0 -L/T
\end{eqnarray}  
This does not work well in our case (Fig.~3): $ln~p$ is not a linear 
function of $\beta=1/T$.  We can discard the assumption that $L$ is
constant and instead use eq.(10) to get an idea of $L$ using values of
$dp/dT, p$ and $T$ from Fig.~1.  If this is done then at 6 MeV 
temperature, the value of $L$ turns out to be 54 MeV and at 7 MeV
temperature this reaches 70 MeV.  Since the binding energy per particle
for an infinite cluster is 16 MeV, these values are clearly
unacceptably high.

Let us ask what went wrong in going from eq.(9) to eq.(11).  The passage
from eq.(10) to eq.(11) assumed that the latent heat is independent
of temperature.  We will show that this is approximately correct.  However,
the approximation $v_{gas}\simeq T/p$ is very inaccurate and depending
upon the temperature, corrections due to $v_{liq}$ can be significant.
When all this is taken into account eq.(9) is satisfied remarkably well.
We elaborate first on the latent heat.  For definiteness
fix on the isothermal at 7 MeV.  From Fig.~1 we
can determine the density at which the system enters the 
purely gaseous phase and its energy per particle 
from $\sum n_ke_k/N$ where $e_k=(3/2)T+e_k(int)$.  Here
$e_k(int)$ consists of volume energy (which is negative),
surface energy and as well contributions from
excited states.  The expressions are given in section II.  The nucleons
are passing from a liquid state (from a composite with $k\simeq 2000$)
to the gaseous phase.  The energy per particle in the liquid phase 
was taken from a composite of 1950 particles.  Reasonable variation
around this number will only change the calculated value
slightly.  Latent heat per particle calculated is 12.66 MeV
at temperature 6 MeV and 11.55 MeV at 7.5 MeV.

By far the major error is in assuming that $v_{gas}\simeq T/p$.  The 
pressure is given by $p=(M/\tilde V)T$ and not $(N/\tilde V)T$ where $M$ is the
multiplicity and $N$ of course the total number of particles. Thus
$p=(M/N)T/v_{gas}$.  The factor $(M/N)$ when the system just turns 
into a pure gas phase is 0.276 at T=6.0 Mev, 0.194 at 7.0 MeV and 0.152
at 7.5 MeV.  Writing $\alpha$ for $M/N$ we find that eq.(10) should
be rewritten as
\begin{eqnarray}
\frac{dp}{dT}=\frac{Lp}{T^2\alpha[1-p/(\rho_0\alpha T)]}
\end{eqnarray}
where we have used the fact that $v_{liq}=1/\rho_0$.  As an example,
at 7 MeV $\frac{dp}{dT}=0.0535$fm$^{-3}$ from Fig.~1 and 0.0530fm$^{-3}$ 
estimated from formula (12).  At 6 MeV the corresponding numbers are
0.0129fm$^{-3}$ and 0.0108fm$^{-3}$ respectively.

\section{The limit large $k_{max}$.}
We will now consider changes in the values of various quantities as
we change from one large value of $k_{max}$ to another large value
of $k_{max}$.  For definiteness we will concentrate on one isothermal,
for example the $T$=7 MeV case.  In Fig.~1 we have two regions: pure
gas phase and the co-existence phase.  The pure gas phase is trivial.
Nothing changes as we go from one large $k_{max}$ to another; $k_{max}$=
2000 is large enough in this case.  It is easy to see why results become
insensitive to changes in the value of $k_{max}$.
In the gas phase there is no population in high $k$ composites so it
does not matter whether the summation stops at a given high value of $k_{max}$
or another high value of $k_{max}$.  The situation is more complicated 
but also more interesting in the co-existence region as we have
population both at the lower end and the higher end of $k$.

In Fig.~4 we plot the values of $\mu$ and pressure calculated for $k_{max}$
in the range $k_{max}$= 2000 to 5000.  This is done at a fixed value
of $\rho/\rho_0=0.3$ which is in the co-existence region.
As shown in the figure, both curves are well
fit by a parametrisation $a+b\exp(-ck_{max})$ where $k_{max}\geq 2000$,
with values given in the caption.
This means that within the accuracy
with which this calculation was carried out, the values of $\mu$ and pressure
in the infinite $k_{max}$ limit are -18.504 MeV and 0.0294 MeV fm$^{-3}$
respectively.

A similar calculation as above was done for $\rho/\rho_0$=0.4.  The fitted
values of $a,b$ and $c$ for $\mu$ were -18.504 MeV, 0.33892 MeV and 0.0003197
respectively.  For pressure, the parameters were 0.0294 MeV fm$^{-3}$,
0.007633 MeV fm$^{-3}$ and 0.0003906 respectively.  Note that the extrapolation
demonstrates that that neither $\mu$ nor the pressure change in the
co-existence region in the limit $k_{max}\rightarrow\infty$.  This firmly
establishes the present model as a model of liquid-gas phase transition
as was stated in section III. 

Lastly we want to estabish a scaling law.  Given the fractional occupation
$f_k(k_{max})=kn_k/N$ for a large value of $k_{max}$, do we know the fractional
occupation $f_{k'}(k'_{max})$ for another large value of $k_{max}$?  Based
on the discussion, so far we expect that if one is 
in the purely gas phase $f_k(k_{max})=f_k(k'_{max})$ and  
this is indeed the case.  

In the co-existence phase a 
lowest order approximation is based on the following approximation.  We
expect the fractional occupation to match near the beginning ($k$ small),
near the end (near $k\leq k_{max}$ and $k'\leq k'_{max}$) and in between
there is almost no occupation.  Thus for $k$ small $f_k(k_{max})\simeq
f_k(k'_{max})$ and near the high end $f_k(k_{max})\simeq f_{k'}(k'_{max})$
where $k_{max}-k=k'_{max}-k'$.  This is not very accurate but
an accurate representation for low $k$ is given using the parametrisation:
\begin{eqnarray}
ln~f_k(k_{max})=ln~f_k(k'_{max})+\frac{k}{T}
b[\exp(-ck_{max})-\exp(-ck'_{max})].
\end{eqnarray}
An equation relating the large clusters can also be written down, but the
functional form is quite complicated.

\section{Small Systems: An Exact Canonical Model Solution}
The model can be solved when the number of particles $N$ is finite.  
Extensive use of the canonical model has been made to fit
experimental data \cite{Dasgupta3} so just an outline will be presented
for completeness.  Among other applications, the canonical model
can be used to study finite particle number effects on phase transition
characteristics.

Consider again $N$ identical particles in an enclosure $V$ and temperature
$T$.  These $N$ nucleons will combine into monomers, dimers, trimers etc.
The partition function of the system in the canonical ensemble
can be written as
\begin{eqnarray}
Q_N=\sum\prod_i\frac{(z_i)^{n_i}}{n_i!}
\end{eqnarray}
Here $z_i$ is the one particle
partition function of a composite which has $i$
nucleons.  We already encountered $z_i$ in section II: 
$z_i=z_i(tran)z_i(int)$
with $z_i(tran)$ and $z_i(int)$ given in detail.
Other forms for $z_i$ can be used in the method outlined
here.  The summation in eq.(14) is over all partitions which satisfy
$N=\sum in_i$.  The summation is non-trivial as the number of partitions which
satisfy the sum is enormous.  We can define a
given allowed partition to be a channel.  The probablity of the occurrence
of a given channel $P(\vec n)\equiv P(n_1,n_2,n_3....)$ is
\begin{eqnarray}
P(\vec n)=\frac{1}{Q_N}\prod\frac{(z_i)^{n_i}}{n_i!}.
\end{eqnarray}
The average number of composites of $i$ nucleons is easily seen from
the above equation to be
\begin{eqnarray}
\langle n_i \rangle = z_i\frac{Q_{N-i}}{Q_N}
\end{eqnarray}
Since $\sum in_i=N$, one readily arrives at a recursion relation
\cite{Chase}
\begin{eqnarray}
Q_N=\frac{1}{N}\sum_{k=1}^{N}kz_kQ_{N-k}
\end{eqnarray}
For one kind of particle, $Q_N$ above is easily evaluated on a computer for
$N$ as large as 3000 in matter of seconds.  It is this recursion relation
that makes the computation so easy in the model.  Of course, once one has
the partition function all relevant thermodynamic quantities can be
computed.  For example, eq. (7) still gives the expression for pressure
although one could correct for the center of mass motion by reducing
the multiplicity by 1: $p=T(M-1)/\tilde V$.  The chemical potential
can be calculated from $\mu=F(N)-F(N-1)$ where the free energy is
$F(N)=-T~ln~Q_N$ which is readily available from the calculation.

In Fig.~5 we show an example of the canonical model calculation.  
The temperature is 6 MeV.  
The number of particles $N$ is 200.  The value of the largest allowed 
cluster $k_{max}$ is also 200.  This would be a typical nuclear
physics case.  In the same figure we also show the result of a grand
canonical calculation with the same $k_{max}$ (of course for the 
grand canonical $N$ is very large).  At small density the results are 
the same but they become different at larger densities.  In the canonical
result there is a small region where $dp/d\rho$ is negative.  This
is a finite particle number effect since for large $N$ (grand canonical
result) any negative compressibilty disappears.  Negative compressibility
can lead to negative value for $c_p$ \cite {Dasgupta3}.  The grand canonical
result shows that for $k_{max}$=200 typical liquid-gas co-existence
is not found and there is no region where $p$ is constant when the density
changes.

\section{Discussion}
Results in section III to section VI show that the model of excited
matter breaking up into clusters with saturation properties
leads to a first order phase transition.  This has relevance
to heavy ion collisions at intermediate energy but may have significance
in other areas of physics as well.  This model for first order
phase transitions is extremely easy to
implement.  A very significant advantage of the model is that it
can be solved not only in the thermodynamic limit (large $N$) but
also for a finite number of particles.  Thus one can study how observables
change as one progresses from small to large systems.

We like to end this discussion by noting that in spite of a very
different approach that is adopted here to arrive at the key
equations 5, 7 and 8, formally the pressure and density equations
have the same structure as those encountered in the well-known
Mayer cluster expansion \cite {Pathria}.  
These are $p=\frac{(2\pi mT)^{3/2}}{h^3}
\sum_{k=1}^{\infty} exp(\beta\mu k)b_k$ and $\rho=\frac{(2\pi mT)^{3/2}}
{h^3}\sum_{k=1}^{\infty}\exp(\beta\mu k)kb_k$.  Here instead of 
the cluster integral $b_k$ we have $k^{3/2}z_k(int)$.

\section{Acknowledgement} 
One of the authors (S.D.G) is indebted to Aram Mekjian for collaboration
over many years in the subject of heavy ion collisions.  Many
of the ideas amplified in this work owe their origin to these collaborations.
The same author is also indebted to Champak Baran Das for past joint
ventures.  This work is supported
in part by the Natural Sciences and Engineering Research Council of Canada
and in part by the Quebec Department of Education.

\begin{figure}
\epsfxsize=5.5in
\epsfysize=7.0in
\centerline{\epsffile{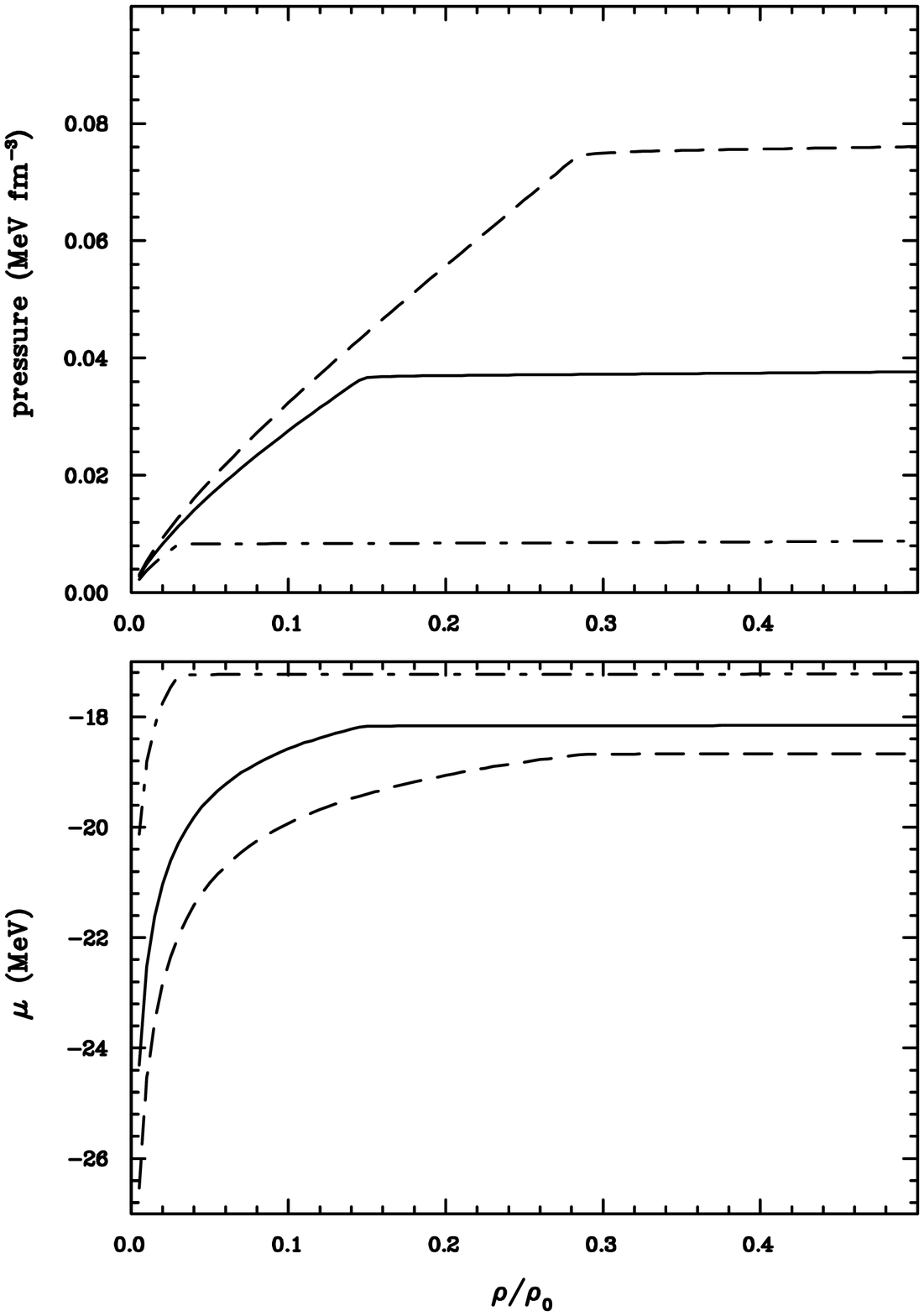}}
\vskip 0.8 true cm
\caption{ Behavior of pressure
$p$ and chemical potential $\mu$ against $\rho/\rho_0$ for 3 different
temperatures : dash (7.5 MeV), solid (7 MeV) and dash-dot (6 MeV). 
We identify as purely gas phase the region where 
the pressure and the chemical potential $\mu$ rise with density
and the co-existence region where they remain constsnt.} 
\end{figure}

\begin{figure}
\epsfxsize=5.5in
\epsfysize=7.0in
\centerline{\epsffile{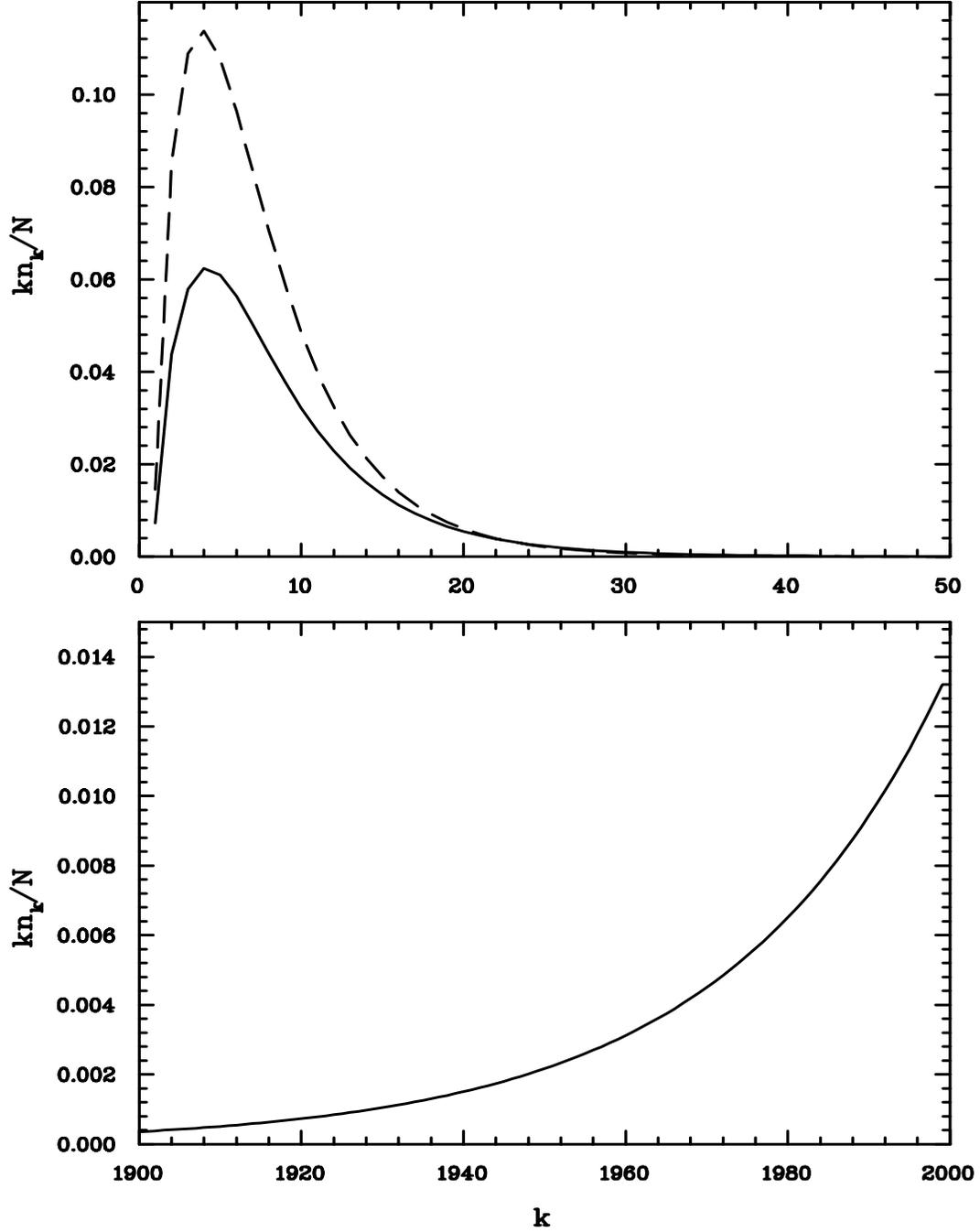}}
\vskip 0.8 true cm
\caption{ At temparature 7 MeV yields of composites at two densities
$\rho/\rho_0$=0.12 (gas phase) and $\rho/\rho_0$=0.22 (the co-existence
phase).  For the first case there are no heavy composites (dashed line).  The
sum rule $\sum kn_k/N=1$ is already satisfied to good accuracy 
by $k=40$.  In the second case (solid line) there are light particles (less
than 50 nucleons) and there are heavy particles (greater than 1800 particles).
Together these exhaust the sum rule.  In $k$ space there is a huge gap
for particles between large and small.  The occupation number in this region
is very close to 0.}
\end{figure}

\begin{figure}
\epsfxsize=4.5in
\epsfysize=6.0in
\centerline{\epsffile{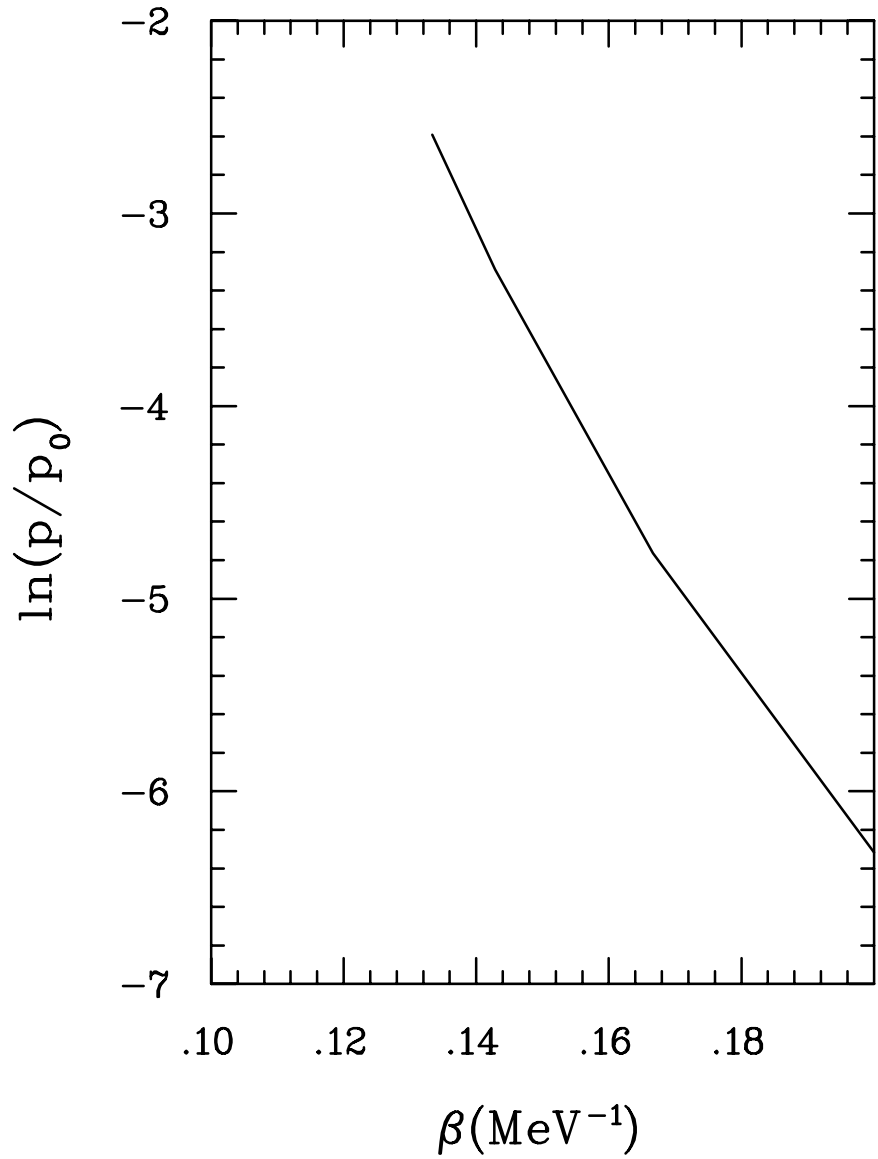}}
\vskip 0.8 true cm
\caption{ A plot of $ln(p)$ against the inverse of temperature.
The relationship is not linear. Here $p_0$ is 1 MeV fm$^{-3}$.}
\end{figure}

\begin{figure}
\epsfxsize=5.5in
\epsfysize=7.0in
\centerline{\epsffile{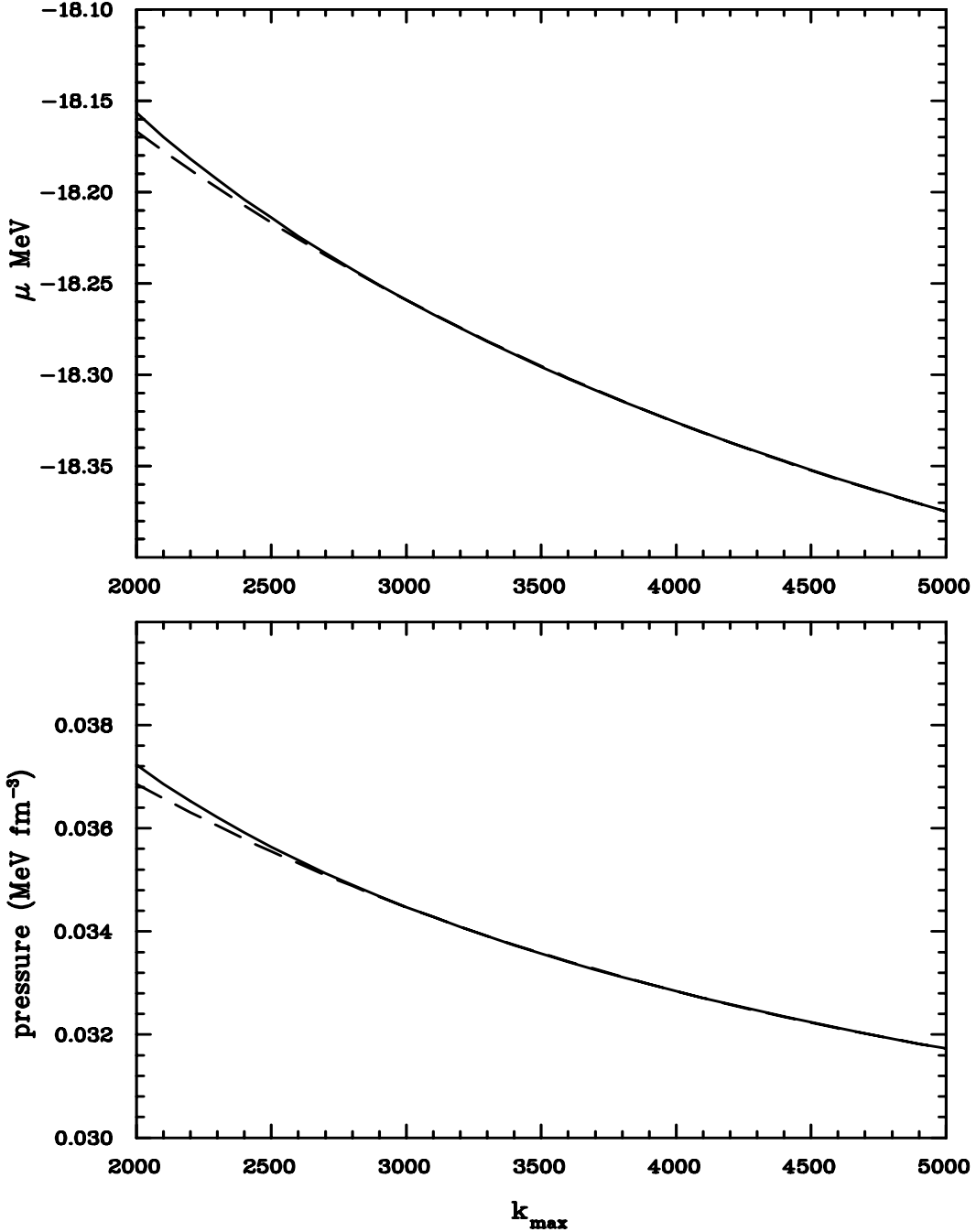}}
\vskip 0.8 true cm
\caption{ The solid curve in the upper panel is a plot of $\mu$ against 
$k_{max}$ in the range of $k_{max}$=2000 to 5000 with density 
at $\rho/\rho_0$=0.3 and temperature 7 MeV.  The dashed curve 
is a fit with the parametrisation $a+b\exp(-ck_{max})$.  The values of the 
fit parameters are $a=-18.504$ MeV, $b$=0.33748 MeV and $c$=0.0003842.
Similar quantities for pressure in the lower panel.
The fit parameters are $a$=0.0294 MeV fm$^{-3}$,
$b$=0.007503 MeV fm$^{-3}$ and $c$=0.0003842.  Similar curves for
$\rho/\rho_0$=0.4 yield equally good fit and
give the same values for $a$ but
different values for $b$ and $c$.}
\end{figure}

\begin{figure}
\epsfxsize=5.5in
\epsfysize=7.0in
\centerline{\epsffile{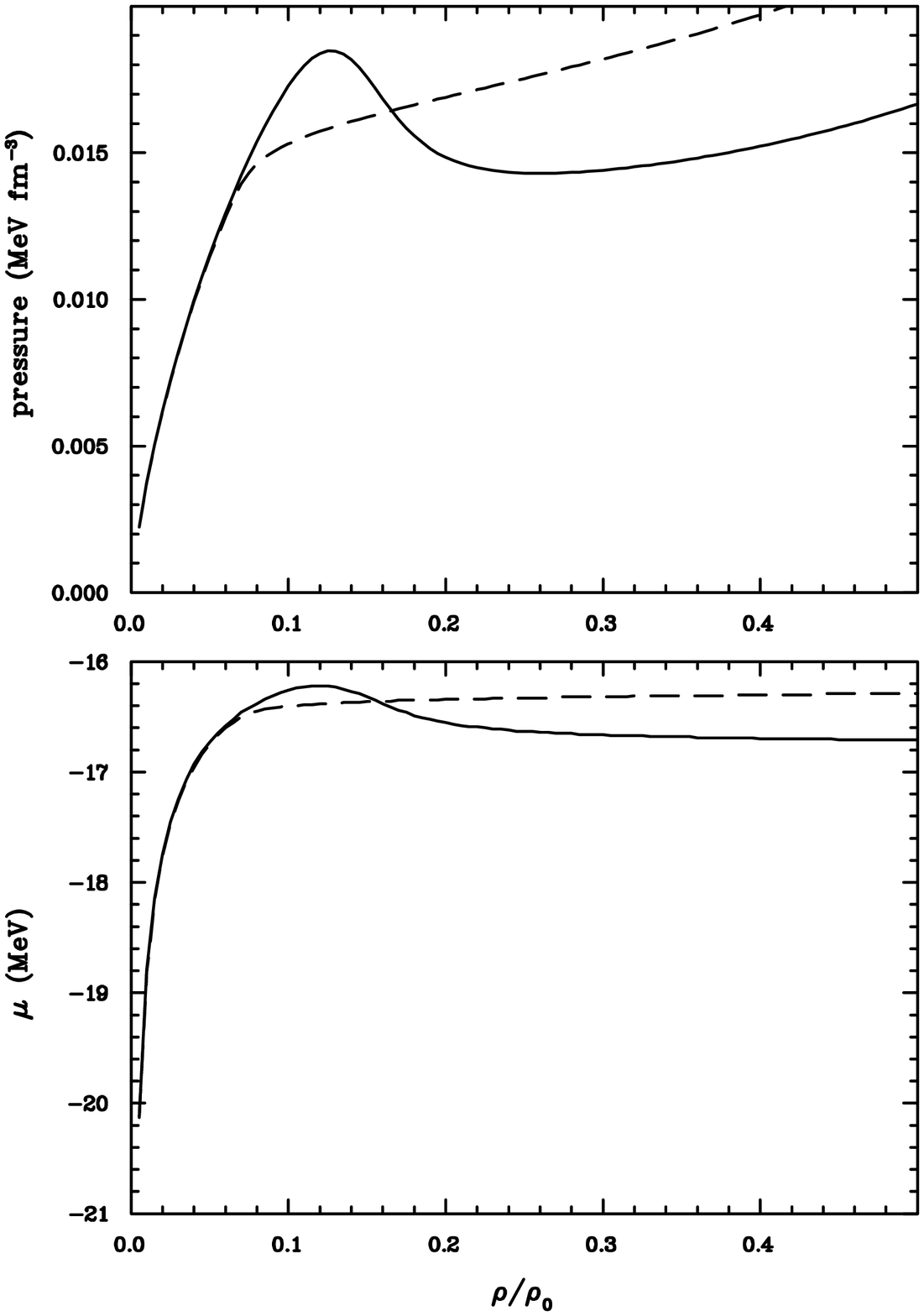}}
\vskip 0.8 true cm
\caption{ The solid curve in the upper panel is a plot of pressure against
density in the canonical model.  The number of particles is $N$=200
exactly and $k_{max}$ is also 200.  Note that there is a region of
negative compressbilty.
The dashed curve is the grand canonical
result with the same $k_{max}$.  The two curves coincide at low
density.  Note that in the grand canonical model the increase of pressure
with density goes down later but never disappears for this low $k_{max}$.
The lower panel compares the chemical potentials.}

\end{figure}


\begin{references}

\bibitem{Dasgupta1} S. Das Gupta, and A. Z. Mekjian, Phys. Rep.{\bf 72},
131 (1981)
 
\bibitem{Dasgupta2} S. Das Gupta and A. Z. Mekjian, Phys. Rev. C{\bf 57},
1361 (1998)

\bibitem{Bugaev} K. A. Bugaev, M. I. Gorenstein, I. N. Mishustin, and
W. Greiner, Phys. Rev C{\bf 62}, 044320 (2000)

\bibitem{Bondorf} J. P. Bondorf, A. S. Botvina, A. S. Iljinov, I. N. 
Mishustin and K. Sneppen, Phys. Rep. {\bf 257}, 133 (1995)

\bibitem{Gross} D. H. Gross, Phys. Rep. {\bf 279}, 119 (1997)
1040 (1995)

\bibitem{Randrup} J. Randrup, and S. E. Koonin, Nucl. Phys. A{\bf 471},
355c (1987)

\bibitem{Reif} F. Reif, {\it Fundamentals of statistical and thermal 
physics}(McGraw Hill, New York, 1965)Chapter 8.

\bibitem{Dasgupta3} C. B. Das, S. Das Gupta, W. G. Lynch, A. Z. Mekjian,
and M. B. Tsang, Phys. Rep. {\bf 406}, 1, (2005)

\bibitem{Chase} K. C. Chase, and A. Z. Mekjian, Phys. Rev C{\bf 52},R2339,
(1995)

\bibitem{Pathria} R. K. Pathria, {\it Statistical Mechanics}(Pergamon
press, Toronto, 1980)Chapter 9.

\end{references}
\end{document}